\documentclass[10pt]{iopart}

\usepackage{soul,color}
\usepackage{iopams}
\expandafter\let\csname equation*\endcsname\relax
\expandafter\let\csname endequation*\endcsname\relax
\usepackage{amsmath}
\usepackage{setspace}
\usepackage{graphicx} 
\usepackage{capt-of} 
\usepackage{amstext}
\usepackage[ngerman,english]{babel}
\usepackage{xcolor}
\usepackage{booktabs}
\usepackage{cite}
\usepackage{hyperref} 
\hypersetup{colorlinks=false, citebordercolor=green}
\definecolor{colorThree}{rgb}{0.61, 0.87, 1.0}
\definecolor{colorTwo}{rgb}{0.99, 0.93, 0.0} 
\definecolor{colorOne}{rgb}{0.5, 1.0, 0.0}

\soulregister\ref{7}
\soulregister\cite{7}
\begin{document}
\title{State engineering of impurities in a lattice by coupling to a Bose Gas}

\author{Kevin Keiler $^1$ and Peter Schmelcher $^{1,2}$}
\address{$^1$Zentrum f\"ur Optische Quantentechnologien, Universit\"at
	Hamburg, Luruper Chaussee 149, 22761 Hamburg, Germany}
\address{$^2$The Hamburg Centre for Ultrafast Imaging, Universit\"at
	Hamburg, Luruper Chaussee 149, 22761 Hamburg, Germany}
\ead{\mailto{kkeiler@physnet.uni-hamburg.de}, \mailto{pschmelc@physnet.uni-hamburg.de}}


\onehalfspacing

\begin{abstract}
We investigate the localization pattern of interacting impurities, which are trapped in a lattice potential and couple to a Bose gas. For small interspecies interaction strengths, the impurities populate the energetically lowest Bloch state or localize separately in different wells with one extra particle being delocalized over all the wells, depending on the lattice depth. In contrast, for large interspecies interaction strengths we find that due to the fractional filling of the lattice and the competition of the repulsive contact interaction between the impurities and the attractive interaction mediated by the Bose gas, the impurities localize either pairwise or completely in a single well. Tuning the lattice depth, the interspecies and intraspecies interaction strength correspondingly allows for a systematic control and engineering of the two localization patterns. The sharpness of the crossover between the two states as well as the broad region of their existence supports the robustness of the engineering. Moreover, we are able to manipulate the ground state's degeneracy in form of triplets, doublets and singlets by implementing different boundary conditions, such as periodic and hard wall boundary conditions.  
\end{abstract}

\newpage
\section{Introduction}
The interest in the properties and dynamics of ultracold atomic mixtures has been substantially increasing in the last few decades. This is not only due to their high degree of controllability, especially of the underlying trapping potentials and inter-atomic interactions \cite{confine,feshbach}, but also because they show a plethora of intriguing phenomena. These range from pair-tunnelling effects in lattices \cite{pflanzer1,pflanzer2} to phase separation processes \cite{phase_sep1,ofir,phase_sep2,phase_sep3} and composite fermionization \cite{cf,pyzh,zollner}, and go as far as the spontaneous generation of solitons \cite{simos1,lia}. Compared to systems with a single species the multi-component case of different bosons \cite{spec_el,wiemanRB} allows for correlations to appear not only within one type of species, but especially between different bosonic species. One-dimensional systems are of particular interest, since they allow for strong correlations in the dilute regime \cite{1d_1,1d_2} which is related to the inverse scaling of the effective interaction strength to the density \cite{1d_eff_corr_1,1d_eff_corr_2}. \\\\
A specific case of bosonic mixtures is given by the immersion of a minority species, consisting of a few particles and typically called impurities, into a majority species of many particles. Such setups have been studied theoretically  \cite{blume_imp,cuc_imp,massignan_review,grusdt,volosniev,zinner_pol,garcia,lampo1,lampo2,knoerzer} and experimentally \cite{corn_pol_sim,trans_imp,naegerl_bloch,catani,fukuhura} for a single impurity, as simulator for polaron physics, as well as for many impurities \cite{zwerger_casimir,fleischhauer_ind,zinner_ind,jie_chen}. Especially the latter case is of immediate interest since the bath, into which the impurities are immersed, mediates an effective attractive interaction between the impurities, leading to a clustering of these very particles \cite{jaksch_pol,jaksch_clus,jaksch_trans}.\\\\
One could think of exploiting this mediated interaction in order to configure the impurities, e.g. in a lattice, in a controlled and systematic manner. Such state preparations are indeed relevant for applications e.g. in quantum information processing and atomtronics \cite{holland,schlagheck,olsh_atomtronics}. One pathway is to use a small number of minority atoms in order to influence a larger number of majority atoms. In a triple-well structure this is achieved by increasing the number of minority atoms in the central well, exploiting the increase of atom-atom interactions, and thereby initiating and enhancing tunnelling from the left to the right well of the majority atoms \cite{bec_transistor}. Pushing this to the extreme, it is in principle possible to implement a single-atom-transistor, which shall serve as a switch \cite{single_atom_transistor1,single_atom_transistor2,fischer1,fischer2,fischer3}. In other atomtronic switching devices, often a triple-well is considered, where one identifies the wells as source, gate and drain. The middle well which is called the gate serves as a mediator of particle transfer between the outer wells, i.e. the source and the drain \cite{zinner_nat_com,foerster}. Such electronic analogues have the potential to serve as building blocks for cold atom-based quantum computation \cite{quantum_gate1,quantum_gate2} as well as atom chip technologies \cite{atom_chip}. A problem most of these systems suffer from, is the fact that they are considered to be isolated and would lose their coherence when coupled to an environment due to dissipation. However, dissipation can also be used as a resource for quantum state engineering \cite{diss_state1,diss_state2,diss_state3}.    \\\\
In the present work, we explicitly take advantage of the coupling to an environment in order to engineer the properties of impurities in a triple-well. Instead of following the path that is mentioned above in the context of atomtronics, we control the behaviour of a minority species by coupling to a majority one.  Hence, we concentrate on static configurations of the impurities, which might be used as an input to logical gates or starting-points for the control of particle transfer. In the system under investigation, the competition of the attractive interaction which is mediated by the majority species and the repulsive contact interaction among the minority impurity species allows for the engineering of the impurity distribution in the lattice. The induced interaction hereby depends on the interplay between the lattice depth and the interspecies interaction strength, whereas the impurity repulsion can be directly influenced by the intraspecies interaction strength among the impurities. Setting the latter to zero and increasing the lattice depth and the interspecies interaction strength, the ground state wave function undergoes a transition from an uncorrelated to a highly correlated state, which manifests itself in the localization of the lattice atoms in the latter regime \cite{keiler}. This means that all impurity atoms cluster in a single well, while the majority Bose gas atoms are expelled from it. In the present work we go a significant step beyond the latter scenario and include a repulsive intraspecies interaction among the impurities. We find that for small interspecies interaction strengths and depending on the lattice depth the system can be divided into the unit filling insulator state and one extra particle delocalized over the wells and the state where all impurities populate the energetically lowest Bloch state. In particular, we show that for large interspecies interaction strengths our binary mixture exhibits two specific impurity distributions for a fractional filling of the lattice. These are either a pairwise clustering of the impurities or a full localization. For periodic boundary conditions the many-body ground states exhibiting these two distributions are threefold degenerate (triplets) in both regimes of pairwise and complete clustering and can be changed to doublets or singlets by implementing different boundary conditions. \\\\
Our work is structured as follows: In section \ref{setup} we present the system under investigation consisting of a bosonic impurity species which is trapped in a one-dimensional lattice with periodic or hard wall boundary conditions and couples to a Bose gas of a second majority species of bosons. Afterwards, we briefly discuss the computational method which is used for obtaining our results. Sections \ref{cs} and \ref{intra} provide a thorough analysis of the system's possible many-body ground states in dependence of the interspecies interaction strength, the lattice depth and the intraspecies interaction strength among the impurities. We conclude in section \ref{bc} with a discussion of the ground state's degeneracy in the different regimes in dependence of the boundary conditions. In section \ref{conclusion} we summarize our findings and present possible applications and future studies.
\section{Setup and methodology}\label{setup}
Our system consists of a mixture of two bosonic species. The bosonic A species is trapped in a one-dimensional lattice with periodic or hard wall boundary conditions. It is immersed in a Bose gas of a second B species of bosons obeying the same boundary conditions but without the lattice potential. This setup lies within reach of current experimental techniques, since beyond controlling the dimensionality, various trapping potentials for the atoms can be achieved, including in particular one-dimensional ring geometries \cite{boshier} and box potentials \cite{box_pot}. The optical lattice potential for the A atoms (impurities) does not affect the Bose gas, which is achievable by choosing the corresponding laser wavelengths and atomic species \cite{spec_sel_lat}. Thereby, we create a two-component system with each species being trapped individually. Furthermore, we introduce a coupling Hamiltonian $\hat{H}_{AB}$ between the two species. Both subsystems are confined to a longitudinal direction, accounting for the one-dimensional character, and excitations in the corresponding transversal direction are energetically suppressed and can therefore be neglected. This finally results in a Hamiltonian of the form $\hat{H}=\hat{H}_A+\hat{H}_B+\hat{H}_{AB}$. 
\begin{figure}[b]
	\centering
	\includegraphics[scale=0.35]{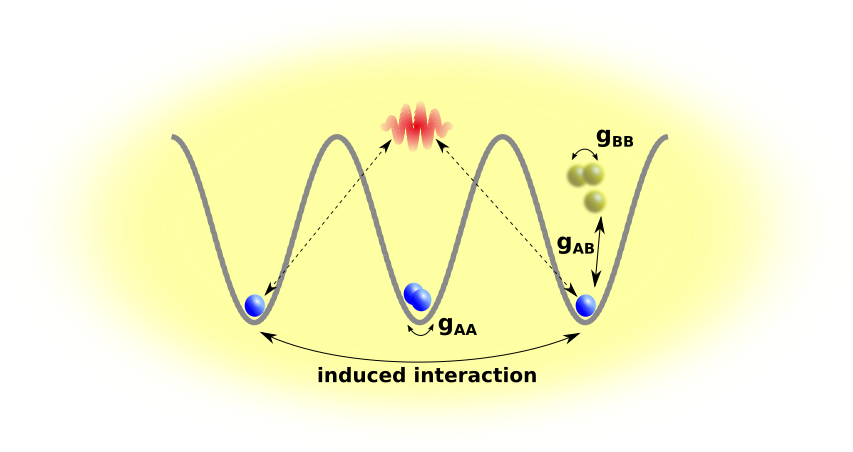}
	\caption{Sketch of the two-component mixture. The impurities interact repulsively via an intraspecies contact interaction of strength $g_{AA}$ and via an interspecies interaction of strength $g_{AB}$ with the atoms of the Bose gas. Due to the latter coupling a Bogoliubov mode (red) in form of a phonon in the Bose gas of species B (yellow) mediates an attractive (long-range) interaction between the impurities (blue).}
	\label{figure2}	
\end{figure}
The Hamiltonian of the A species reads

\begin{equation}
\hat{H}_A=\int_{0}^{L} \text{dx} \, \hat{\chi}^{\dagger}(\text{x}) \Big [ -\frac{\hbar^{2}}{2 m_A} \frac{\text{d}^{2}}{\text{dx}^{2}} + V_0 \sin^{2}\Big(\frac{\pi k \text{x}}{L}\Big)+ g_{AA} \; \hat{\chi}^{\dagger}(\text{x}) \hat{\chi}(\text{x}) \Big ] \hat{\chi}(\text{x}),
\end{equation}
where $\hat{\chi}^{\dagger}$ is the field operator of the lattice A bosons, $m_A$ their mass, $V_0$ the lattice depth, $g_{AA}$ the intraspecies interaction strength, $k$ the number of wells in the lattice and $L$ is the length of the system.
The B species is described by the Hamiltonian of the Lieb-Liniger model \cite{LL1,LL2,LLnagerl} for periodic boundary conditions

\begin{equation}
	\hat{H}_B=\int_{0}^{L} \text{dx} \; \hat{\phi}^{\dagger}(\text{x}) \Big [ -\frac{\hbar^{2}}{2 m_B} \frac{\text{d}^{2}}{\text{dx}^{2}} + g_{BB} \; \hat{\phi}^{\dagger}(\text{x})\hat{\phi}(\text{x}) \Big ] \hat{\phi}(\text{x}),
\end{equation}
where $\hat{\phi}^{\dagger}$ is the field operator of the B species, $g_{BB}>0$ is the interaction strength of the two-body contact interaction among the B atoms and $m_B$ is the corresponding mass.  Moreover, we assume equal masses for the species $m_A=m_B$. The interaction between the species A and B is given by
\begin{equation}
	\hat{H}_{AB}= g_{AB} \int_{0}^{L} \text{dx} \; \hat{\chi}^{\dagger}(\text{x}) \hat{\chi}(\text{x}) \hat{\phi}^{\dagger}(\text{x}) \hat{\phi}(\text{x}),
\end{equation}
where $g_{AB}$ is the interspecies interaction strength. The interaction strengths $g_{\alpha}$ ($\alpha\in\{A,B,AB\}$) can be expressed in terms of three dimensional s-wave scattering lengths $a^{3D}_{\alpha}$, when assuming the above-mentioned strong transversal confinement with the same trapping frequencies $\omega^{\sigma}_{\perp}=\omega_{\perp}$ for both species $\sigma \in \{A,B\}$. In this case it is possible to integrate out frozen degrees of freedom, leading to a quasi one-dimensional model with $g_{\alpha}=2\hbar\omega_{\perp}a^{3D}_{\alpha}$.\par
Throughout this work we consider a triple-well and focus on the scenario of small particle numbers with four impurities $N_A=4$, thereby having fractional filling in the lattice, and $N_B=10$ atoms in the Bose gas. The interaction among the latter atoms is set to a value where the depletion is negligible in case of no interspecies coupling, i.e. $g_{BB}/E_R \lambda=6.8 \times 10^{-3}$, with $E_R=(2\pi\hbar)^{2}/2m_A \lambda^{2}$ being the recoil energy and $ \lambda=2L/k$ the optical lattice wavelength. \par
Our numerical simulations are performed using the \textit{ab-initio} Multi-Layer Multi-Configuration Time-Dependent Hartree method for bosonic (fermionic) Mixtures (ML-MCTDHX) \cite{mlb1,mlb2,mlx}, which is able to take all correlations into account. Within ML-MCTDHX one has access to the complete many-body wave function which allows us consequently to derive all relevant characteristics of the underlying system. In particular, this means that we are able to characterize the system in terms of number states by projecting onto an appropriate basis \cite{ns_analysis1,ns_analysis2}. Besides investigating the quantum dynamics it allows us to calculate the ground (or excited) states by using either imaginary time propagation or improved relaxation \cite{meyer_improved}, thereby being able to uncover also possible degeneracies of the many-body states. In standard approaches for solving the time-dependent Schr{\"o}dinger equation, one typically constructs the wave function as a superposition of time-independent Fock states with time-dependent coefficients. Instead, the ML-MCTDHX approach considers a co-moving time-dependent basis on different layers, i.e. the Fock states and thus the single particle functions spanning them are time-dependent, in addition to time-dependent coefficients. This leads to a significantly smaller amount of basis states that are needed to obtain an accurate description and eventually reduces the computation time.
\section{State control and engineering}
Let us analyze the ground state of our mixture in dependence of the lattice depth $V_0$, the interspecies coupling strength $g_{AB}$ and the intraspecies interaction strength $g_{AA}$. As a first step, we calculate the ground state using ML-MCTDHX, thereby obtaining the full wave function. In order to be able to interpret the wave function, we project in a second step the numerically obtained ground state wave function onto number states $|\vec{n}^{A}\rangle\otimes|\vec{n}^{B}\rangle$. The number states $|\vec{n}^{A}\rangle$ for the A species are spanned by generalized Wannier states \cite{kivelson1,kivelson2}, whereas the number states for the Bose gas of species B are either plane waves or infinite square well eigenstates, i.e. the eigenstates of the kinetic energy operator using either periodic or hard wall boundary conditions. As a result, we gain a clear insight into the ground state in the different regimes, which will be defined by the distribution of the A species atoms among the Wannier states or Bloch states. In the following, tensor products  $|\vec{n}^{A}\rangle\otimes|\vec{n}^{B}\rangle$ with different number states $|\vec{n}^{A}\rangle$ will be called configurations. Additionally, we explore the effect of hard wall boundary conditions compared to periodic ones, thereby revealing how they affect the ground state properties. 

\subsection{Localization pattern of impurities}\label{cs}
In the following, we explore the ground state of the system with periodic boundary conditions for varying $V_0$ and $g_{AB}$, fixing the intraspecies interaction strength to $g_{AA}/E_R \lambda=0.0236$. It turns out, that this interaction strength lies within the range of possible values which lead to the manifestation of regimes in which the ground states differ substantially. An analysis of the dependence on $g_{AA}$ is performed in section \ref{intra}. In order to extract information out of the complete many-body wave function, we project onto the above-mentioned number states $|\vec{n}^{A}\rangle\otimes|\vec{n}^{B}\rangle$ and determine the probability of being in the number state $|\vec{n}^{A}\rangle$ for the impurity A species, irrespective of the number state configurations of the B species, namely
\begin{equation}
P(|\vec{n}^{A}\rangle)= \sum_{i} |\langle \vec{n}^{B}_i |\otimes\langle \vec{n}^{A}|\Psi\rangle|^{2},
\end{equation}
where $\{|\vec{n}^{B}_i\rangle\}$ could be any number state basis set of the Bose gas with fixed particle number and $|\Psi\rangle$ is the total many-body ground state wave function. 
In order to associate the impurity state $|n^A_1,n^A_2,n^A_3\rangle$ with a spatial distribution we construct the number states either with a generalized Wannier basis of the lowest band or the corresponding Bloch basis set, indicated in the following by the subscript $W$ or $Bl$, respectively. In principle it is necessary to consider also Wannier states of higher bands, but it turns out that the many-body ground state wave function is approximately well described by a superposition of tensor products of number states $|\vec{n}^{A}\rangle\otimes|\vec{n}^{B}\rangle$ with the Fock space of the A species restricted to the lowest band. Number states spanned by Wannier or Bloch states of higher bands do not contribute. The order of the entries in the number state, built of Wannier states, is connected to the localization of the Wannier states in the wells from left to right, e.g. $n_1$ describes the number of A atoms in the left localized Wannier state of the lowest band. In contrast to that, the ordering for the number states, built of Bloch states, follows the energy of the Bloch states of the lowest band, e.g. $n_1$ corresponds to the energetically lowest Bloch state. The transformation between Bloch and Wannier states is given by 
\begin{equation}
w^{b}_R(\text{x})=\frac{1}{\sqrt{k}}\sum_{p}\exp(-ipR)\phi^{b}_p(\text{x}),
\label{def_wannier}
\end{equation}
where $w_R(\text{x})$ is the Wannier state associated with the position $R$ of the corresponding well, $k$ the number of lattice sites, $p$ the momentum of the Bloch states $\phi_p(\text{x})$ and $b$ the index of the band. \par
\begin{figure}[t]
	\centering
	\includegraphics[scale=0.3]{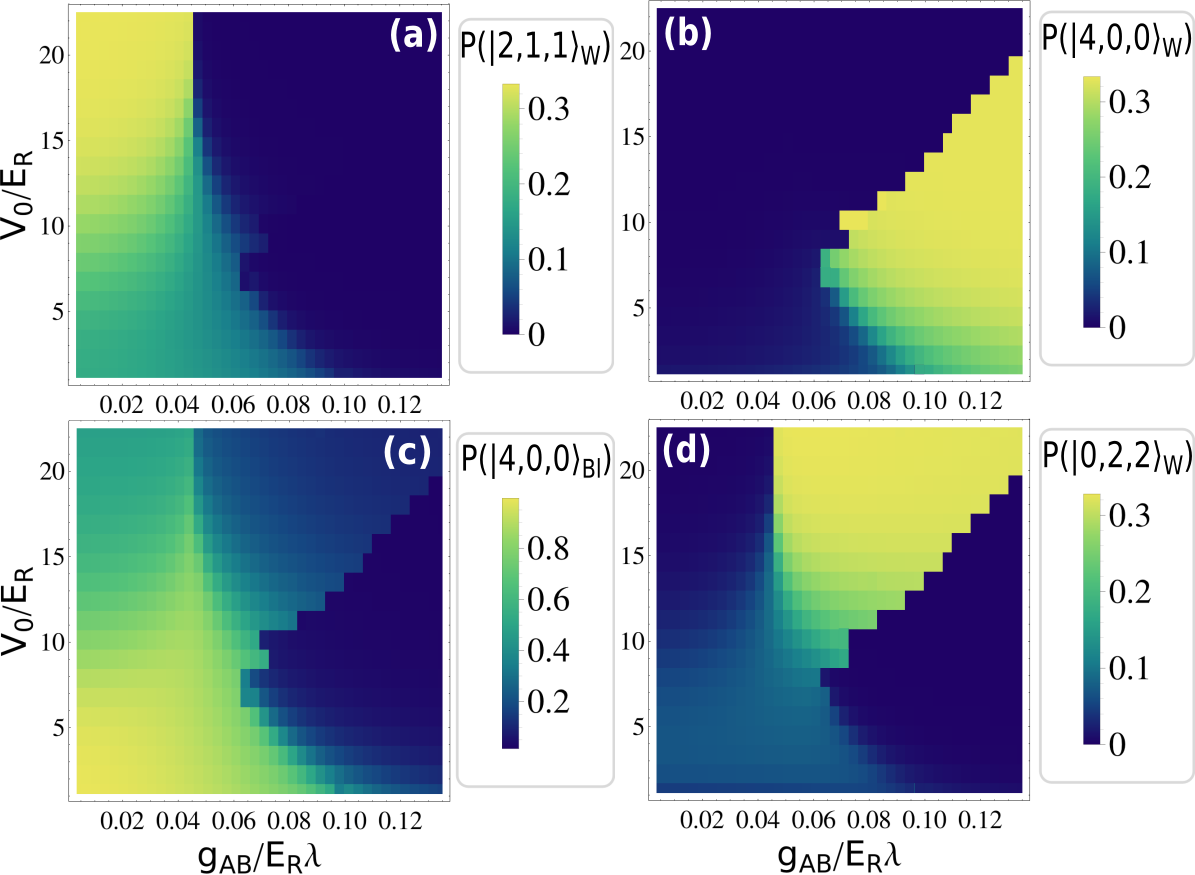}
	\caption{Probability of finding the impurity A species atoms (a) being localized separately in different wells with the one extra particle being delocalized over all the wells, (b) all localized in a single well, (c) residing in the energetically lowest Bloch state and (d) localized pairwise in two different wells in dependence of the lattice depth $V_0$ and the interspecies interaction strength $g_{AB}$ for $g_{AA}/E_R \lambda=0.0236$. The number states which are spanned by Wannier states [(a),(b),(d)] are chosen exemplarily since all wells are equally valid due to translational symmetry, and therefore the corresponding translationally equivalent configurations show the same behaviour.}
	\label{figure1a_1d}	
\end{figure} 
Figure \ref{figure1a_1d} shows that the many-body ground state can be divided into four different regions. For small lattice depths and interspecies interaction strengths all four particles of the A species populate the energetically lowest Bloch state [figure \ref{figure1a_1d}(c)]. Increasing the lattice depth for small $g_{AB}$ three lattice atoms will localize separately in different wells with the one extra particle being delocalized over all the wells [figure \ref{figure1a_1d}(a)]. This can be seen by the fact that the corresponding probability is given by $P(|2,1,1\rangle_{W})=\frac{1}{3}$, meaning that the remainder of the probability is equally distributed over the states $|1,2,1\rangle$ and  $|1,1,2\rangle$, resulting in the state
\begin{equation}
|\Psi\rangle_{M}=\frac{1}{\sqrt{3}}\Big [|2,1,1\rangle_{W}\otimes|\psi^{1}_B\rangle+|1,2,1\rangle_{W}\otimes|\psi_B^{2}\rangle+|1,1,2\rangle_{W}\otimes|\psi_B^{3}\rangle \Big].\;
\label{eq:psi_M}	
\end{equation}
Due to translational symmetry each of the states contributes equally with a probability of $\frac{1}{3}$.
The reader should note that the structure of the wave function, given in equation \ref{eq:psi_M}, can only be uncovered by the procedure of projection onto number states $|\vec{n}^{A}\rangle\otimes|\vec{n}^{B}\rangle$ and is not explicitly given by the numerical simulation. Therefore, $|\Psi\rangle_{M}$ is not the exact result of ML-MCTDHX but an approximation to it. Our approach is to perform the correlation-including ML-MCTDHX calculations and to subsequently analyze them. \par
Interestingly, the increase of $g_{AB}$ leads to two different regions in the configuration space, in contrast to the case of $g_{AA}=0$ \cite{keiler}. For $g_{AA}=0$, increasing the lattice depth and the interspecies interaction strength, the ground state wave function undergoes a transition from an uncorrelated to a highly correlated state, which manifests itself in the localization of the lattice atoms in the latter regime of large lattice depths and interspecies interaction strengths. This means that all A atoms cluster in a single well, while the B atoms are expelled from it. This clustering can be understood in terms of an attractive induced impurity-impurity interaction, which is mediated by the B species (cf. figure \ref{figure2}). In contrast to that, allowing for a repulsive intraspecies interaction of strength $g_{AA}$ between the A atoms counteracts the induced interaction. However, these two types of interactions do not simply add up, since the induced interaction is of long-range type, whereas the intraspecies interaction is of contact type. As a result, depending on the choice of the lattice depth $V_0$ and the interspecies interaction strength $g_{AB}$ the impurities of species A either accumulate all in one well [figure \ref{figure1a_1d}(b)], which is already happening in the case of $g_{AA}=0$, or pairwise in two different wells [figure \ref{figure1a_1d}(d)]. Apparently, in the case of pairwise localization of the lattice atoms [figure \ref{figure1a_1d}(d)] the repulsive interaction counteracts an accumulation of all lattice atoms due to the induced interaction.
Performing a thorough analysis of the excitation spectrum of the many-body system, we find that the ground state for large $g_{AB}$ is in both regions threefold degenerate (triplet). This essentially means that in the corresponding region the ground states are either given by 
\begin{equation}
	 |2,2,0\rangle_{W}\otimes|\bar{\Psi}^{1}_B\rangle\text{,}\; |2,0,2\rangle_{W}\otimes|\bar{\Psi}^{2}_B\rangle \; \text{and} \; |0,2,2\rangle_{W}\otimes|\bar{\Psi}^{3}_B\rangle \quad \text{or by} \;
	\label{eq:deg_220}
\end{equation}

\begin{equation}
|4,0,0\rangle_{W}\otimes|\Psi^{1}_B\rangle\text{,}\; |0,4,0\rangle_{W}\otimes|\Psi^{2}_B\rangle \; \text{and} \; |0,0,4\rangle_{W}\otimes|\Psi^{3}_B\rangle.
\label{eq:deg_400}
\end{equation}
The states $\{|\Psi^{i}_B\rangle\}$ and $\{|\bar{\Psi}^{i}_B\rangle\}$ \cite{comment_psi} are each normalized to unity and incorporate the localization effect of the A species by e.g. spatially avoiding the impurities correspondingly, which can be seen in the one-body density (cf. section \ref{bc}, figure \ref{figure5a_5f}).
Due to the degeneracy of the ground state one can choose such superpositions of the states in equation \ref{eq:deg_220} and \ref{eq:deg_400}, which preserve the translational invariance of the total Hamiltonian. For this reason the probabilities in figures \ref{figure1a_1d} (a),(b),(d) are bounded by a maximum value of $1/3$ \cite{comment_imag}. 
Furthermore, it is now possible to use this degeneracy in order to select any of the states in the respective degenerate manifold. Technically, this is simply done by applying a small asymmetry to the lattice potential, thereby energetically favouring one of the above-mentioned states. For example, marginally increasing the depth of the left well of the lattice potential will break the translational symmetry and thereby lift the degeneracy, such that the ground state is solely given by $|4,0,0\rangle_{W}\otimes|\Psi^{1}_B\rangle$ in one regime. In this sense, with the lattice depth of individual wells we have introduced an additional control parameter for the manipulation of impurity configurations.\par
So far, we gained insight into the state configurations that are populated by the impurities. While this finding itself allows for a systematic control of the impurities in the lattice, it is nevertheless of interest in which way the correlation with the Bose gas impacts this very species.
Therefore, we also analyze the probability distribution $P(|\vec{n}^{B}\rangle)$ of the number states $|\vec{n}^{B}\rangle$ that build the corresponding B species states $\{|\Psi^{i}_B\rangle\}$ and $\{|\bar{\Psi}^{i}_B\rangle\}$. We find that for each of the strongly coupled degenerate ground states the B species states $\{|\Psi^{i}_B\rangle\}$ and $\{|\bar{\Psi}^{i}_B\rangle\}$ each populate the same number states with equal probability, i.e. $|\langle \vec{n}^{B}|\Psi^{i}_B\rangle|^{2}=|\langle \vec{n}^{B}|\Psi^{j}_B\rangle|^{2}$ with $i,j\in\{1,2,3\}$, while differing only in a relative phase for each coefficient of the number state. This is why it is sufficient to show only the probability distribution of one B species state for the states in equations \ref{eq:deg_220} and \ref{eq:deg_400}.
As a basis for the number states $|n^B_1,n^B_2,n^B_3,n^B_4,n^B_5\rangle$ we choose plane waves $(1/\sqrt{L})\exp(i\kappa\text{x})$, with wave vectors $\kappa=2\pi z/L$ ($z=0,\pm1,\pm2,...$), where $n_1$ corresponds to the $\kappa=0$ mode, $n_2, n_3$ correspond to $z=\pm1$ and $n_4, n_5$ to $z=\pm2$. It turns out that the population of all higher momentum states is negligible, which we checked explicitly by projecting onto the corresponding number states.
\begin{figure}[t]
	\centering
	\includegraphics[scale=0.3]{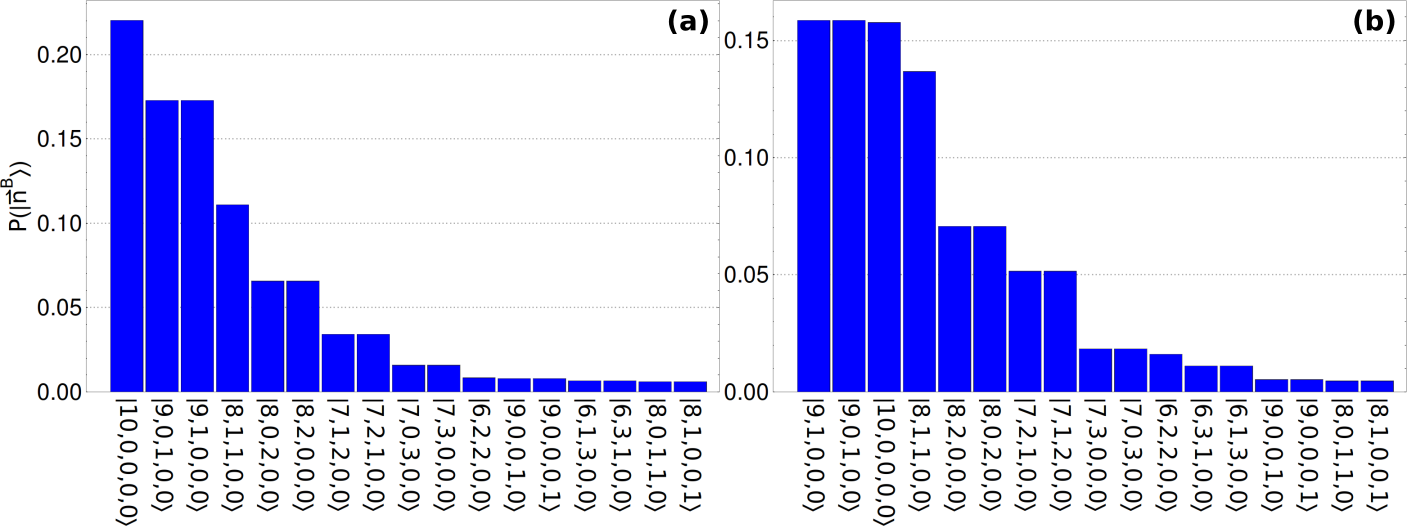}
	\caption{Probability distribution for the states of the B species for $g_{AA}/E_R \lambda=0.0236$, $g_{AB}/E_R \lambda =0.135$ and (a) $V_0/E_R=18$ or (b) $V_0/E_R=22.5$. (a) describes ground states with a complete localization of A atoms in a single well (cf. equation \ref{eq:deg_400}), whereas (b) describes ground states with a pairwise localization in two wells (cf. equation \ref{eq:deg_220}).}
	\label{figure3a_3b}	
\end{figure}
In figure \ref{figure3a_3b}, we see that due to the correlation with the impurity species the Bose gas can no longer be described by a single number state with all particles occupying the $\kappa=0$ mode \cite{comment_gBB}. Interestingly, it is also not sufficient to consider only single particle excitations. It is rather necessary to consider up to four particle excitations for the B species states in both regimes. Comparing the B species states in the two regimes, one finds that they strongly populate the same number states, but differ w.r.t. the quantitative distribution among those number states. For example, in figure \ref{figure3a_3b}(a) the number states $|9,1,0,0,0\rangle$ and $|9,0,1,0,0\rangle$ are less populated than the number state $|10,0,0,0,0\rangle$, whereas their probability exceeds that of $|10,0,0,0,0\rangle$ in figure \ref{figure3a_3b}(b).
These findings support the choice of our treatment of the many-body problem using a method that is in particular capable of taking all necessary correlations into account. An approximation of the many-body Hamiltonian which relies on few-particle excitations, will not capture the localization pattern presented in figure \ref{figure1a_1d}.
\subsection{Dependence on the intraspecies coupling}\label{intra}
In the previous subsection, we have identified four different number state configurations for the impurity species, while assuming a fixed intraspecies interaction strength of $g_{AA}/E_R \lambda =0.0236$. However, it is not clear whether this holds for a broad regime of couplings $g_{AA}$. In order to explore the range of validity of this crossover, we fix the lattice depth as well as the interspecies interaction strength such that for $g_{AA}=0$ we arrive at a degenerate subspace of ground states given by equation \ref{eq:deg_400}, instead of the one given by equation \ref{eq:deg_220} for $g_{AA}/E_R \lambda=0.0236$. The reader should note that for $g_{AA}=0$ the correlated region, which is split into two sub-regions for $g_{AA}/E_R \lambda=0.0236$, is solely described by the degenerate manifold in equation \ref{eq:deg_400} (see \cite{keiler}). Again, we calculate the probability $P(|\vec{n}^{A}\rangle)$, while varying the coupling strength $g_{AA}$ for fixed $V_0$ and $g_{AB}$. 
\begin{figure}[b]
	\centering
	\includegraphics[scale=0.4]{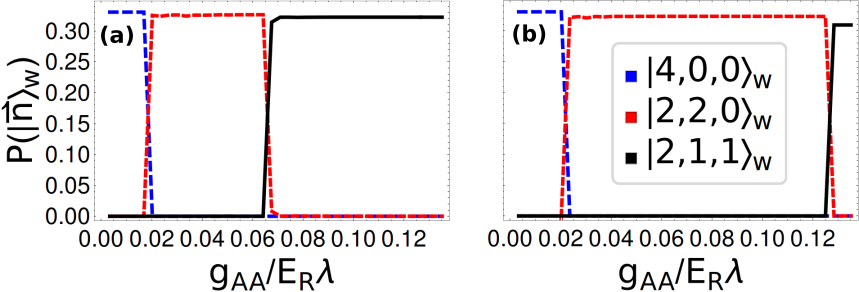}
	\caption{Probability distribution of the number state configurations of the impurities in dependence of the intraspecies coupling strength $g_{AA}$ for $V_0/E_R=22.5$  and (a) $g_{AB}/E_R \lambda=0.084$ or (b)  $g_{AB}/E_R \lambda=0.135$.}
	\label{figure4a_4b}	
\end{figure} 
In figure \ref{figure4a_4b}, we see that for small $g_{AA}$ the correlated region is well described by a single triplet of ground states, as in the case of $g_{AA}=0$, where all impurities accumulate in a single well. A further increase of the intraspecies interaction strength leads to a break-up of the cluster into two pairs (equation \ref{eq:deg_220}) and finally results in a ground state with all A atoms localized separately in different wells with the one extra particle being delocalized over all the wells (cf. equation \ref{eq:psi_M}). Essentially, this means that one needs a certain intraspecies interaction strength $g_{AA}$ between the impurities in order to arrive at a crossover diagram as in figure \ref{figure1a_1d}. Below that critical value the crossover is well captured by the $g_{AA}=0$ case, including, if $g_{AA}\neq0$, a regime for small $g_{AB}$ and large $V_0$ where the ground state is given by $|\Psi\rangle_M$ (equation \ref{eq:psi_M}). In other words, for small $g_{AA}$ the crossover diagram in figure \ref{figure1a_1d} will consist of three different regimes, where the two regimes in figure \ref{figure1a_1d}(b) and (d) will merge into a single one, describing complete localization of the impurities in a single well (equation \ref{eq:deg_400}). Thus, the ground state comprising pairwise localization of the A atoms will not exist in this case.
\par Qualitatively, the crossover in figure \ref{figure4a_4b} can be understood again in terms of a competition between the attractive induced interaction and the repulsive intraspecies contact interaction. For small $g_{AA}$ the induced interaction is dominating the behaviour of the A species atoms, leading to their complete localization in a single well. At a certain interaction strength $g_{AA}$ this is no longer the case, resulting in a pairwise accumulation of A atoms. Apparently, the different nature of the competing interactions (long-range and contact) does not lead to a trivial reduction of the four-impurity cluster to a three-impurity cluster, but rather leaves this out as a possibility and directly favours a two-impurity cluster. 
Astonishingly, the crossover between the different configurations is very sharp, such that the system occupies only one of the triplets without superposing them. Furthermore, it is possible to control the width of the plateau of the pairwise localization by adjusting the interspecies coupling strength $g_{AB}$ correspondingly. The plateau (red) corresponding to a degenerate manifold of ground states with pairwise localization of A atoms is much broader for a larger value of $g_{AB}$ [cf. figure \ref{figure4a_4b}(b)]. This also means that a smaller value of the interspecies interaction strength $g_{AB}$ leads to a smaller critical value of $g_{AA}$ [cf. figure \ref{figure4a_4b}(a)] at which the transition to the ground state $|\Psi\rangle_M$ takes place (black plateau, equation \ref{eq:psi_M}).
In essence, we find that by tuning the intraspecies interaction strength, we are able to control and engineer the localization of the A atoms in the lattice. The sharpness of the crossovers allows for a clear and systematic way of choosing the ground states, while the broad plateaus make the triplets robust with respect to fluctuations of $g_{AA}$. In this sense, the coupled system serves as a transistor-like switching device for number state preparation of impurities in a lattice.

\subsection{Boundary conditions}\label{bc}
Our findings in the previous sections so far relied on the fact that we assumed periodic boundary conditions. It is therefore of immediate interest in which way the localization pattern in figure \ref{figure1a_1d}(b) and (d) depends on the choice of the boundary conditions. For this reason, we consider solely values of the lattice depth $V_0$ and interspecies interaction strength $g_{AB}$ of the crossover diagram such that we arrive at the ground state configurations in equations \ref{eq:deg_220} and \ref{eq:deg_400}. The corresponding values are given in table \ref{table_states}. Subsequently, for those two regimes we change the boundary conditions to hard wall boundary conditions. Obviously, this change will break the translational symmetry of the Hamiltonian. Instead, the Hamiltonian now obeys parity symmetry, suggesting that the former ground state degeneracy of a triplet might now be given by a doublet. Indeed, for the states comprising complete localization of the A atoms in a single well (equation \ref{eq:deg_400}) we arrive at the subspace of degenerate ground states, where the atoms of species A localize in the outer wells, i.e. $|4,0,0\rangle_{W}\otimes|\Psi^{1}_B\rangle$ and $|0,0,4\rangle_{W}\otimes|\Psi^{2}_B\rangle$ \cite{comment_Wannier}. However, the degenerate ground state with pairwise localization for periodic boundary conditions does not exhibit any degeneracy for hard wall boundary conditions anymore. In this sense, the ground state in that regime is given by a non-degenerate parity symmetric ground state of the form $|2,0,2\rangle_{W}\otimes|\bar{\Psi}_B\rangle$ (singlet).\par
Investigating the one-body density of the Bose gas of species B, one gets an intuition for the reason why in one region the ground state triplet becomes a doublet and in the other it becomes a singlet for hard wall boundary conditions. The one-body density shows that the Bose gas accumulates in the centre of the box potential \cite{comment_box} (not shown here), because it is energetically favourable for the majority of the bosons of the Bose gas to occupy the energetically lowest eigenmode of the box potential and not accumulate to either side of the box. As a consequence, the A atoms localize in the outer wells due to the repulsive coupling to the B species, thereby avoiding occupation of the middle well. For the ground states comprising complete localization of A atoms in a single well, the restriction of the A atoms to the outer wells allows for two possible many-body states out of three in equation \ref{eq:deg_400}, whereas for the ground states with pairwise localization of A atoms only a single state in equation \ref{eq:deg_220} obeys this restriction. \par
\begin{figure}[t]
	\centering
	\includegraphics[scale=0.35]{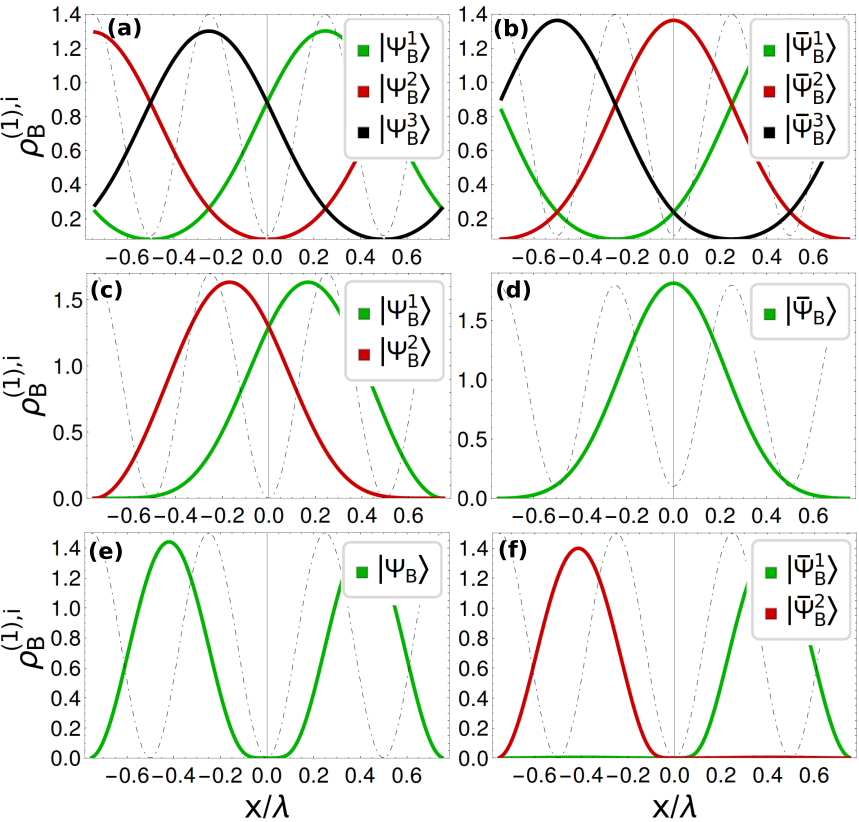}
	\caption{One-body density $\rho^{(1),i}_B$ for the states of the B species in table \ref{table_states} with $g_{AA}/E_R \lambda=0.0236$, $g_{AB}/E_R \lambda=0.135$ and (left column) $V_0/E_R=18$ or (right column) $V_0/E_R=22.5$. The first row is based on periodic boundary conditions, the second row is based on hard wall boundary conditions and the third one incorporates a repulsive Gaussian potential for the B species in the centre of the box in addition to the hard wall boundary conditions. The dotted line is a sketch of the lattice potential, indicating the position of the wells \cite{comment_Wannier}. }
	\label{figure5a_5f}	
\end{figure}
Following the above line of argumentation, one might now ask whether it is possible to change the singlet ground state into a doublet and vice versa by forcing the Bose gas out of the centre and thereby to either side of the box. In order to achieve this we implement a repulsive Gaussian potential for the B species in the middle of the box which is localized in the middle well of the lattice and has an amplitude $A_0$ that is approximately twice as large as $V_0$, i.e. $A_0/E_R=50$. It turns out that this procedure has the effect we aim for, leaving an imprint on the one-body density of the corresponding B species states [cf. figure \ref{figure5a_5f} (e) and (f)], which is defined as
\begin{equation}
\rho_B^{(1),i}(\text{x}) = \int \text{dx}_2...\text{dx}_{N_B} |\Psi^{i}_B(\text{x},\text{x}_2,...,\text{x}_{N_{B}})|^{2},
\label{eq_B_dens}
\end{equation}
integrating over all B atoms except for one.\\
In figure \ref{figure5a_5f}, we show the one-body density $\rho^{(1),i}_B$ ($i\in\{1,2,3\}$) of the states of the B species, following the nomenclature in table \ref{table_states}. Figure \ref{figure5a_5f}(e) shows that the one-body density of the B species indeed exhibits a minimum in the centre of the box, leading to an accumulation to both sides. Consequently, the A atoms accumulate in the middle well, such that only one many-body ground state out of equation \ref{eq:deg_400} fulfills this restriction, namely $|0,4,0\rangle_{W}\otimes|\Psi_B\rangle$. In contrast to that, for pairwise impurity localization forcing the B species out of the centre of the box allows for two possible ground states from equation \ref{eq:deg_220} [cf. figure \ref{figure5a_5f}(f)], namely $|2,2,0\rangle_{W}\otimes|\bar{\Psi}^{1}_B\rangle$ and $|0,2,2\rangle_{W}\otimes|\bar{\Psi}^{2}_B\rangle$. Table \ref{table_states} summarizes the engineering of the degenerate subspace of the ground state in dependence of the boundary conditions for pairwise and complete localization of the A atoms in a single well.
\begin{table}[t]
	\centering
	\captionof{table}{Degenerate subspaces of the ground state for different boundary conditions for $g_{AB}/E_R \lambda=0.135$ and $g_{AA}/E_R \lambda=0.0236$. The third row incorporates a repulsive Gaussian potential for the B species in the centre of the box in addition to the hard wall boundary conditions. The different values of the lattice depth $V_0$ lead either to a complete localization of A atoms in a single well or to a pairwise localization \cite{comment_Wannier}.}
	\label{table_states}
	\begin{tabular}{ccc}
		\br
		boundary conditions  & $V_0/E_R=18$ & $V_0/E_R=22.5$ \\ \mr
		periodic & $|4,0,0\rangle_{W}\otimes|\Psi^{1}_B\rangle$  & $|2,2,0\rangle_{W}\otimes|\bar{\Psi}^{1}_B\rangle $ \\
		& $|0,4,0\rangle_{W}\otimes|\Psi^{2}_B\rangle $ & $|2,0,2\rangle_{W}\otimes|\bar{\Psi}^{2}_B\rangle $ \\
		&$|0,0,4\rangle_{W}\otimes|\Psi^{3}_B\rangle$ & $ |0,2,2\rangle_{W}\otimes|\bar{\Psi}^{3}_B\rangle$ \\\\
		hard walls & $|4,0,0\rangle_{W}\otimes|\Psi^{1}_B\rangle$ & $|2,0,2\rangle_{W}\otimes|\bar{\Psi}_B\rangle$\\
		& $|0,0,4\rangle_{W}\otimes|\Psi^{2}_B\rangle$& \\\\
		hard walls and repulsive Gaussian potential & $|0,4,0\rangle_{W}\otimes|\Psi_B\rangle$ & $|2,2,0\rangle_{W}\otimes|\bar{\Psi}^{1}_B\rangle$\\
		& & $|0,2,2\rangle_{W}\otimes|\bar{\Psi}^{2}_B\rangle$ \\
		\br
	\end{tabular}
\end{table} 
Figure  \ref{figure5a_5f}(a) resembles the case of $g_{AA}=0$, where the particles of species B are expelled from the well where all the A atoms fully localize, leaving an imprint on the one-body density. In figure  \ref{figure5a_5f}(b) the B species atoms need to be expelled from two wells due to the pairwise localization of A atoms. Because of the fact that there are less impurities per well the hereby reduced interspecies interaction allows for a larger one-body density of species B in the region of the pairwise occupied wells. Figure  \ref{figure5a_5f} (c) and (d) are similar to (a) and (b) except for a shifting of the density closer to the centre of the box potential. This is simply an effect of the change to hard wall boundary conditions. Apart from two specific ground state configurations (figure \ref{figure5a_5f}(b) red and \ref{figure5a_5f}(d) green) it is possible to identify any of the many-body ground states in table \ref{table_states} just by analyzing the position of the one-body density of the B species states with respect to the lattice potential - irrespective of the boundary conditions. \par
Thus, we are able to engineer the character of degeneracy of the ground state by choosing the boundary conditions correspondingly (in combination with a Gaussian potential). Combining this with the fact that it is possible to switch between differently localized configurations of the impurities by tuning $V_0$, $g_{AB}$ and $g_{AA}$, one might think of an (adiabatic) particle transfer of the following type. Initially, we prepare the ground state in one of the doublet states (using hard wall boundary conditions), e.g. $|4,0,0\rangle_{W}\otimes|\Psi^{1}_B\rangle$. Increasing the intraspecies interaction strength $g_{AA}$ adiabatically the ground state will reconfigure to the singlet $|2,0,2\rangle_{W}\otimes|\bar{\Psi}_B\rangle$. Essentially, this can be interpreted as a transfer of two impurities from the left to the right well.
\section{Conclusions}\label{conclusion}
We have shown that it is possible to manipulate the configuration space of lattice trapped impurities with fractional filling immersed in a Bose gas. For small interspecies interaction strengths, the impurities populate the energetically lowest Bloch state or localize separately in different wells with the one extra particle being delocalized over all the wells, depending on the lattice depth. In contrast, for large interspecies interaction strengths and depending on the lattice depth and intraspecies coupling we find that the impurities either localize pairwise or completely in a single well of the lattice. Astonishingly, in dependence of the intraspecies and interspecies coupling as well as the lattice depth the system switches between those two internal state configurations, allowing for an engineering of the impurity distribution in a systematic and controlled manner. Furthermore, the change  from periodic to hard wall boundary conditions will reconfigure the ground state from a triplet to either a doublet for ground states where the A atoms fully localize in one well, or to a singlet for ground states where they localize pairwise in one well. We can exploit this degeneracy even further in order to select individual states out of the manifold by applying a small asymmetry to the lattice potential. Eventually, we are not only able to let the impurities cluster in a certain way, but also manipulate in which wells they accumulate. Additionally, we are able to influence the ground state's character of degeneracy. In the spirit of atomtronics, we have developed a switching device for many-body state preparation, thereby controlling the accumulation of impurities in a lattice. This analysis is also applicable for a larger number of particles in the environment, while still remaining in the few particle regime (we have tested this for $N_B\in[10,30]$), resulting in the same crossover to the two localization patterns for large $g_{AB}$. However, such a particle increase will also increase the attractive induced interaction for a given choice of $V_0$ and $g_{AB}$, thereby shifting the transition region. Increasing the number of impurities, the impurities might form multi-atom clusters of different types depending on the parameter regime due to the long-range character of the induced interaction. We have additionally performed calculations for three impurities in the same setup. In this case the splitting into the two regions of pairwise and complete localization for large $g_{AB}$ does not occur. Instead, for large interspecies interaction strengths we find only one regime in which the impurities cluster completely in a single well. Furthermore, it is of interest for future studies how the number of lattice sites affects the localization pattern. A natural question appearing in the case of four lattice sites is related to how the two-impurity clusters will distribute in the lattice. They might appear next to each other or such that one lattice site is empty between two two-impurity clusters. \par
The control over the impurity distribution serves as a perfect starting point for dynamical particle transfer scenarios, where one would prepare the impurities in a fully localized state and dynamically transfer it to a pairwise localized state, thereby transferring two particles. In particular, this could also be of interest for the implementation of quantum logical gates, where the position of the impurities shall indicate a logical operation. Especially the dynamical response may pave new pathways to applications in atomtronics. Moreover, one could think of exploiting this engineering of the impurity distribution in order to create multi-atom clusters which are not of binary type. These clusters could serve as a static-disordered potential for the study of Anderson localization using ultracold atoms as proposed in previous works \cite{castin1,castin2,schneble}.
\ack
The authors appreciate fruitful and insightful discussions with K. Sengstock. K. K. acknowledges helpful discussions with J. Schurer and M. Pyzh. P. S. gratefully acknowledges funding by the Deutsche Forschungsgemeinschaft in the framework of the SFB 925 "Light induced dynamics and control of correlated quantum systems" and support by the excellence cluster "The Hamburg Centre for Ultrafast Imaging-Structure, Dynamics and Control of Matter at the Atomic Scale" of the Deutsche Forschungsgemeinschaft. K. K. acknowledges a scholarship of the Studienstiftung des deutschen Volkes.


\begin{thebibliography}{80}
\bibitem{comment_psi}
In the following, $\{|\Psi^{i}_B\rangle\}$ will refer to the B species states corresponding to fully localized impurity states and $\{|\bar{\Psi}^{i}_B\rangle\}$ corresponding to pairwise localized impurity states.
\bibitem{comment_imag}
However, this effect is solely due to the fact that imaginary time propagation with ML-MCTDHX for obtaining the ground state wave function will conserve translational symmetry, if it is, as we did, initially assumed.
\bibitem{comment_gBB}
Of course, the depletion of the Bose gas due to intraspecies interactions of strength $g_{BB}$ will also lead to population of $\kappa\neq0$ momentum states. But since the depletion is rather small for $g_{AB}=0$ this effect plays a minor role.
\bibitem{comment_Wannier}
The reader should note that the Wannier states building the number states are calculated using hard wall boundary conditions. Furthermore, the states of the B species will change correspondingly. We omit indication of the boundary conditions in order to avoid confusion.
\bibitem{comment_box}
We introduce the equally valid term 'box potential' as a synonym for hard wall boundary conditions.
\end{thebibliography}
\end{document}